\documentclass[conference, a4paper]{IEEEtran}

\IEEEoverridecommandlockouts
\newcommand{\F}{\mathbb{F}}
\newcommand{\Q}{{\cal Q}}
\newcommand{\R}{{\cal R}}
\newcommand{\C}{{\cal C}}

\usepackage{graphicx} % Required for including images
\usepackage{support-caption}
\usepackage{caption}
\usepackage{subcaption}
\usepackage{booktabs} % Top and bottom rules for table
\usepackage[font=small,labelfont=bf]{caption} % Required for specifying captions to tables and figures
\usepackage{amsfonts, amsmath, amsthm, amssymb} % For math fonts, symbols and environments
\usepackage{tikz}
\usepackage{pgfplots}
\pgfplotsset{compat=1.16}
\usepackage{tikzscale}
\usetikzlibrary{calc}
\tikzset{circ/.style={draw,circle,inner sep=0.1pt},topath/.style={to path={|-(\tikztotarget)}}}
\usetikzlibrary{plotmarks}
\usetikzlibrary{arrows, matrix, shapes, decorations, patterns, positioning}

\newtheorem{theorem}{Theorem}
\newtheorem{definition}{Definition}
\newtheorem{lemma}{Lemma}

\newtheorem*{example*}{Example}

\newcommand{\up}[1][i]{\ensuremath^{(#1)}}

\title{On Skew Convolutional and Trellis Codes}

\author{
	\IEEEauthorblockN{Vladimir Sidorenko\IEEEauthorrefmark{1}, Wenhui Li\IEEEauthorrefmark{2}, Onur  G\"unl\"u\IEEEauthorrefmark{3}, and Gerhard Kramer\IEEEauthorrefmark{1}}
	
	\medskip
	\IEEEauthorblockA{\IEEEauthorrefmark{1}Institute for Communications Engineering,
		Technical University of Munich, Germany}
	\IEEEauthorblockA{\IEEEauthorrefmark{2}Skolkovo Institute of Science and Technology, Moscow, Russia}
	\IEEEauthorblockA{\IEEEauthorrefmark{3}Information Theory and Applications Chair,
		Technical University of Berlin, Germany\\
		Emails: vladimir.sidorenko@tum.de, w.li@skoltech.ru, guenlue@tu-berlin.de, gerhard.kramer@tum.de}
	\thanks{
		\emph{ V. Sidorenko} is on leave from the Institute for Information
		Transmission Problems, Russian Academy of Sciences.
		 His
		work was supported by the European Research Council (ERC) under
		the European Union's Horizon 2020 research and innovation
		programme (grant agreement No. 801434) and by the Institute for Communications
		Engineering at the Technical University of Munich. 	  
	      }
	\thanks{ The work of \emph{ W. Li} was supported by RFBR, project No. 20-07-00652.}
	\thanks{The work of \emph{ O. G\"unl\"u} was supported by the German Federal Ministry of Education and Research (BMBF) within the national initiative for ``Post Shannon Communication (NewCom)'' under Grant 16KIS1004.}
	\thanks{	The work of \emph{ G. Kramer} was supported in part by the German Research Foundation (DFG) through  Grant KR 3517/9-1. }
}

\begin{document}
\maketitle

\begin{abstract}
	Two new classes of skew codes over a finite field $\F$ are proposed, called skew convolutional codes and skew trellis codes.  These two classes are defined by, respectively, left or right  sub-modules over the skew fields of fractions of skew polynomials over $\F$. The skew convolutional codes  can be represented as periodic time-varying ordinary convolutional codes. The skew trellis codes are in general nonlinear over $\F$. Every code from both classes has a code trellis and can be decoded by  Viterbi or BCJR algorithms. 
\end{abstract}

\section{Introduction}
	
Convolutional codes were introduced by Elias in 1955 \cite{Eli}. These codes became popular when in 1967 Viterbi  invented his decoding algorithm \cite{Vit} and Forney \cite{For}  drew a \emph{code trellis} which made understanding the Viterbi algorithm easy and its maximum-likelihood nature obvious. 
Convolutional codes are widely used in telecommunications, e.g., in Turbo codes and in the WiFi IEEE 802.11 standard, in cryptography, etc. 

The most common versions are \emph{binary} convolutional codes;
\emph{non-binary} convolutional codes are used for higher orders of modulation~\cite{Ouahada} or data streaming  \cite{Holzbaur}.
It is known that periodic time-varying convolutional codes improve the free distance and weight distribution  over fixed codes, see; e.g.,   Mooser \cite{Moo} and  Lee \cite{Lee}. This is a motivation to introduce the new class of linear \emph{skew convolutional codes} that can be represented as ordinary periodic non-binary  convolutional codes. The new class is defined as a \emph{left} module over a skew field $\Q$ that will be introduced later. A \emph{right} module over $\Q$ defines another interesting class of \emph{nonlinear trellis codes}.

Our goal is to define and to give a first encounter with the introduced skew codes. The proofs and additional results about the skew convolutional codes as well as more examples and references can be found in the journal version \cite{SLGK2020}.

\section{Skew convolutional codes}

\subsection{Skew polynomials and fractions}

 Consider a  field $\F$ and an automorphism
$\theta$ of the field. 
Later on, we will use the finite field $\F=\F_{q^m}$ with the Frobenius
automorphism
\begin{equation}\label{theta}
	\theta(a)=a^{q}
\end{equation} 
for all $a\in \F$. Denote by $\R = \F[D;\theta]$ the noncommutative
ring of skew polynomials in $D$   over $\F$  (with zero derivation)
\begin{equation*}
\resizebox{1\hsize}{!}{$\F[D;\theta]=\{a(D)= a_0 +a_1D +\dots + a_{n}D^{n} \ |\ a_i\in \F \mbox{ and } n\in \mathbb{N}\}.$}
\end{equation*}
The addition in $\R$ is as usual.
The multiplication is defined by the basic rule 
$$Da=\theta(a)D$$
and is extended to all elements of $\R$ by
associativity and distributivity.
The ring $\R$ has a unique left skew field  of fractions $\Q$, from which it inherits its linear algebra properties; see,  e.g., \cite{clark:2012} for more details and \cite{SLGK2020} for some examples.

\subsection{Definition of skew convolutional codes}

Much of linear algebra can be generalized from vector spaces over a  field to (either left or right) modules over the skew field $\Q$.  Indeed, it is shown in \cite[Theorem 1.4]{clark:2012} that  any left $\Q$-module $\C$ is free, i.e., it has a basis,
and any two bases of $\C$ have the same cardinality, which is the dimension of $\C$.

\begin{definition}[Skew convolutional code]\label{def:SkewCode}
A skew convolutional $[n,k]$ code $\C$ over the field $\F$  is a left sub-module of dimension $k$ of the free module $\Q^n$.
\end{definition}
The elements of the code $\C$ are called its \emph{codewords}.  A codeword is  an $n$-tuple over $\Q$, where every component is a fraction of skew polynomials from $\R$.  The (Hamming) weight of a fraction is the number of nonzero coefficients in its expansion as a left skew Laurent series $\F((D))$ in increasing powers of $D$. The code $\C$ is $\F=\F_{q^m}$-linear. The \emph{free distance} $d_f$ of a skew convolutional code is defined to be the minimum nonzero weight over all codewords.  

\subsection{Relations with ordinary convolutional codes}
\begin{lemma}
    The class of skew convolutional codes includes ordinary time-invariant (fixed) convolutional codes.
\end{lemma} 
Indeed, when  $\theta = id$, a skew convolutional code coincides with an ordinary convolutional code.

A \emph{generator matrix} of a skew convolutional $[n,k]$  code $\C$ is a $k\times n$ matrix  $G(D)$ over the skew field $\Q$ whose rows form a basis for the code $\C$. If the  matrix $G(D)$ is over the ring $\R$ of skew polynomials, then $G(D)$ is called a \emph{polynomial generator matrix} for $\C$. Every skew code $\C$ has a polynomial generator matrix. Indeed, given a generator matrix $G(D)$ over the skew field of fractions $\Q$, a polynomial generator matrix can be obtained by left multiplying each row of $G(D)$ by the left least common  multiple of the denominators in that row.

\section{Encoding} 

From Definition~\ref{def:SkewCode}, every codeword $v(D)$ of a skew code $\C$, which is an $n$-tuple over the skew field of fractions $\Q$,
\begin{equation}\label{eq:v(d)}
v(D)=\left( v\up[1](D),  \dots,  v\up[n](D)  \right), \   v\up[j](D) \in \Q \ \ \ \forall j,
\end{equation}
can be written as 
\begin{equation}\label{eq:enc1}
v(D) = u(D) G(D), 
\end{equation}
where $u(D)$ is a $k$-tuple ($k$-word) over $\Q$: 
\begin{equation}\label{eq:u(d)}
u(D)=\left( u\up[1](D),  \dots,  u\up[k](D)  \right), \  u\up[i](D) \in \Q \ \ \  \forall i
\end{equation}
and is called an information word, and $G(D)$ is a $k \times n$ generator matrix of $\C$. Relation (\ref{eq:enc1}) already provides an encoder. This encoder  is  an encoder of a block code  over $\Q$ and the skew code $\C$ can be considered as the set of $n$-tuples $v(D)$ over $\Q$ that satisfy (\ref{eq:enc1}), i.e., we have $\C =\{v(D)\}$. 

We write the components of $u(D)$ and $v(D)$ as skew Laurent series 
\begin{equation}\label{eq:u^i}
u\up[i](D) = u\up[i]_0 + u\up[i]_1 D + u\up[i]_2 D^2 + \dots, \  i= 1,\dots, k
\end{equation}
and 
\begin{equation}\label{eq:v^j}
v\up[j](D) = v\up[i]_0 + v\up[i]_1 D +  u\up[i]_2 D^2 + \dots, \ j= 1,\dots, n.
\end{equation}
Actually, in a Laurent series, the lower (time) index of coefficients can be a negative integer, but in practice,  information sequence $u\up[i](D)$ should be causal for every component $i$, that is, the coefficients $u\up[i]_t D^t$ are zeros for time $t < 0$. Causal information sequences should be encoded into causal code sequences, otherwise an encoder can not be implemented, since it would then have to output code symbols before it receives an information symbol.  

Denote the block of information symbols that enters an encoder at time $t=0,1,\dots$ by 
\begin{equation}\label{eq:u_t}
u_t = \left(u_t\up[1],u_t\up[2],\dots u_t\up[k] \right) \in \F^k. 
\end{equation} 
The block of code symbols that leaves the encoder at time $t=0,1,\dots$ is denoted by 
\begin{equation}\label{eq:v_t}
v_t = \left(v_t\up[1],v_t\up[2],\dots v_t\up[n] \right) \in \F^n. 
\end{equation} 
Combining (\ref{eq:u(d)}), (\ref{eq:u^i}), and (\ref{eq:u_t}) we obtain the following information series with vector coefficients  
\begin{equation}\label{eq:u(D)ser}
u(D) = u_0 + u_1 D  + ...+u_t D^t+\dots, \ u(D)\in \F((D))^k.
\end{equation}
Using (\ref{eq:v(d)}), (\ref{eq:v^j}), and (\ref{eq:v_t})  we write  a codeword as a series  
\begin{equation}\label{eq:v(D)ser}
v(D) = v_0 + v_1 D  + ...+v_t D^t+\dots, \  v(D)\in \F((D))^n.
\end{equation}

 We can write a skew polynomial generator matrix $G(D) = \left(g_{ij}(D)\right) \in \R^{k\times n}$ as a skew polynomial with matrix coefficients: 
\begin{equation}\label{eq:G(D)Ser}
G(D) = G_0 + G_1 D + G_2 D^2 + ...+G_\mu D^\mu,
\end{equation}
where $\mu$ is the maximum degree of polynomials $g_{ij}(D)$. Matrices $G_i$ are $k\times n$ matrices over the field $\F$ and $\mu$ is called the generator matrix \emph{memory}.

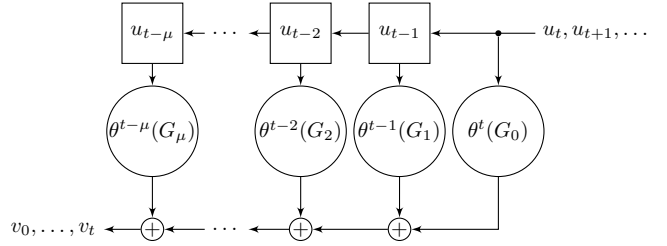
\begin{figure}
	\centering
	\begin{tikzpicture}[scale=0.65, every node/.style={scale=.8}]
	% big circles
	\node(w0) [draw,circle,minimum size=1.5cm,inner sep=0pt] at (0,2) {$\theta^{t-\mu}(G_\mu)$};
	\node(w3) [draw,circle,minimum size=1.5cm,inner sep=0pt] at (3,2) {$\theta^{t-2}(G_2)$};
	\node(w5) [draw,circle,minimum size=1.5cm,inner sep=0pt] at (5,2) {$\theta^{t-1}(G_1)$};
	\node(w7) [draw,circle,minimum size=1.5cm,inner sep=0pt] at (7,2) {$\theta^{t}(G_0)$};
	% squares
	\draw (0,4)  node[minimum size=1cm,draw](s0) {$u_{t-\mu}$};
	\draw (3,4)  node[minimum size=1cm,draw](s3) {$u_{t-2}$};
	\draw (5,4)  node[minimum size=1cm,draw](s5) {$u_{t-1}$};
	% addition,arrows
	\foreach \x in {0,3,5} {
		\draw (\x,0) node(c\x) [circ]{$+$};
		\draw [->,-latex'] (s\x) edge (w\x) (w\x) edge (c\x);
	}
	% u,v part
	\filldraw (7, 4) node(dot) [circle,fill,inner sep=1pt]{};
	\node(u) at (9,4) {$u_t, u_{t+1}, \dots$};
	\draw [->,-latex'] (u) edge (s5);
	\node(v) at (-2,0) {$v_0, \dots, v_t$};
	\draw [->,-latex'] (c0) edge (v);
	% arrows
	\node(udots) at (1.5,4) {$\dots$};
	\node(vdots) at (1.5,0) {$\dots$};
	\draw [->,-latex'] (s5) edge (s3) (s3) edge (udots) (udots) edge (s0);
	\draw [->,-latex'] (c5) edge (c3) (c3) edge (vdots) (vdots) edge (c0);
	\draw [->,-latex'] (dot) edge (w7) (w7) edge [topath](c5);
	\end{tikzpicture}
	\caption{Encoder of a skew convolutional code.}\label{fig:encoder1}
\end{figure}

From (\ref{eq:enc1}), (\ref{eq:u(D)ser}) and (\ref{eq:v(D)ser}) we obtain that $v_t$ is a coefficient in the product of skew series $u(D)$ and skew polynomial $G(D)$, which is the following \emph{skew convolution} (see Fig.~\ref{fig:encoder1})
\begin{equation}\label{eq:enc_conv}
v_t = u_t \theta^t (G_0) + u_{t-1} \theta^{t-1} (G_1) + \dots +  u_{t-\mu} \theta^{t-\mu} (G_\mu),
\end{equation}
where $u_t = 0$ for $t<0$. 
This encoding rule  explains the title \emph{skew convolutional code}, which can be also seen as the set $\C = \{v(D)\}$ of series $v(D)$ defined in (\ref{eq:v(D)ser}).

At time $t$, the decoder receives an information block $u_t$ of $k$ symbols from $\F$ and puts out the code block $v_t$ of $n$ code symbols from $\F$ using (\ref{eq:enc_conv}), hence, the \emph{code rate} is $R=k/n$. The encoder (\ref{eq:enc_conv}) uses  $u_t$ and also $\mu$ previous information blocks $u_{t-1}, u_{t-2}, \dots, u_{t-\mu}$, which should be stored in the encoder's memory. This is why $\mu$ is also the encoder \emph{ memory}. 

The coefficients $\theta^{t-i} (G_i)$, $i=0,1,\dots,\mu$, in the encoder  (\ref{eq:enc_conv}) depend on the time $t$. Hence, the \emph{skew convolutional code is a time varying ordinary convolutional code}. Denote 
\begin{equation}\label{eq:period}
\tau = \min\left\{i>0\ :\ \theta^i(G_j) = G_j \ \ \forall j=0,1,\dots, \mu     \right\}.
\end{equation}
 For the field $\F =\F_{q^m}$  we have $\theta^m = \theta$, hence, the coefficients in (\ref{eq:enc_conv})  are periodic with period  $\tau\le m$, and
the \emph{skew  convolutional code is periodic with period  $\tau \le m$}. If $\tau < m$ then coefficients of polynomials $g_{ij}(D)$ in the matrix $G(D)$ belong to a subfield  $\F_{q^\tau}\subset\F_{q^m}$, and hence $\tau|m$.

The input of the encoder can also be written as an information sequence $u$ of $k$-blocks \eqref{eq:u_t} over $\F$
\vskip -10pt
\begin{equation}\label{eq:u}
u = u_0,\:  u_1, \: u_2 ,  \dots, u_t, \dots \ ,
\end{equation} 
and the output as a code sequence $v$ of $n$-blocks \eqref{eq:v_t} over $\F$
\begin{equation}\label{eq:v}
v = v_0,\:  v_1, \: v_2 ,  \dots,\: v_t ,\: \dots \ .
\end{equation} 
Then, the encoding rule (\ref{eq:enc_conv}) can be written in a scalar form 
\begin{equation}\label{eq:enc_scalar}
v = u G 
\end{equation}
with semi-infinite scalar generator block matrix $G=$
\begin{equation}\label{eq:G}
\left(
\begin{array}{ccccccc}
G_0 & G_1          &  G_2         &  \dots & G_\mu &                           &      \\
	& \theta(G_0)  & \theta(G_1)  & \dots  &       &    \theta(G_\mu)          &           \\
	&		       & \theta^2(G_0)& \dots  &       &    \theta^2(G_{\mu-1})    &\theta^2(G_{\mu}) \\
	&              &              & \dots    \\
\end{array}
\right)
\end{equation}
Thus, a skew convolutional code can be equivalently represented in scalar form as the set $\C=\{v\}$ of sequences $v$ defined in (\ref{eq:v}) that satisfy (\ref{eq:enc_scalar}). By changing  variables $G_i = \theta^i(\widetilde G_i)$ for $i=1,2,\dots,\mu$ we obtain the following result.

\begin{lemma}\label{lem:G_equiv}
	A scalar generator matrix (\ref{eq:G}) can be written in the following equivalent form 
\begin{equation}\label{eq:G_equiv}
G=\left(
\begin{array}{ccccccc}
\widetilde G_0& \theta(\widetilde G_1)&      &\theta^\mu(\widetilde G_\mu)    &                                    &      \\
              & \theta(\widetilde G_0)&\vdots&\theta^\mu(\widetilde G_{\mu-1})&\theta^{\mu+1}(\widetilde G_\mu)    &       \\
              &		                  &      &\vdots                          &\theta^{\mu+1}(\widetilde G_{\mu-1})& \vdots \\
              &                       &      &\theta^\mu(\widetilde G_0)      &\vdots                   \\
              &                       &      &                                &\theta^{\mu+1}(\widetilde G_{0})             \\
\end{array}
\right).
\end{equation}	
\end{lemma}
In case of identity  automorphism, i.e., $\theta = id$, the scalar generator matrix \eqref{eq:G} of the skew code becomes a generator matrix of a fixed convolutional code \cite{JZ}. 

For fixed convolutional codes, polynomial generator matrices with $G_0$ of full rank $k$ are  
of particular interest \cite[Chapter~3]{JZ}. The skew convolutional codes  use the following nice property: if $G_0$ has full rank, then $\theta^i(G_0)$ has full rank as well for all $i=1,2,\dots$. 

Thus, above we proved the following theorem.
\begin{theorem}
	Given a field $\F=\F_{q^m}$ with automorphism $\theta$ in (\ref{theta}), any skew convolutional $[n,k]$ code $\C$ over $\F$ is equivalent to a periodic  time-varying (ordinary) convolutional $[n,k]$ code over $\F$, with period $\tau\le m$  (\ref{eq:period}). If $G(D)$ is a skew polynomial generator matrix (\ref{eq:G(D)Ser}) of the code $\C$, then the scalar generator matrix $G$ of the time-varying code is given by (\ref{eq:G}) or (\ref{eq:G_equiv}). 
\end{theorem}

\section{An example}\label{sec:example}
As an example consider [2,1] skew convolutional code $\C$ over the field $\F_Q = \F_{q^m}=\F_{2^2}$ with automorphism  $\theta(a) = a^q=a^2$, $a\in \F_{2^2}$. The field $\F_{2^2}$ consists of elements $\{0,1,\alpha,\alpha^2\}$, where a primitive element $\alpha$  satisfies $\alpha^2+\alpha+1=0$ and we have the following relations

\medskip
	\begin{tabular}{lc}
%		\hline		
		$\alpha^2 = \alpha +1$,  \\
		$\alpha^3 = 1$,\\
%		\hline
		$\alpha^4 = \alpha$,  \\
%		\hline
	\end{tabular}
\ and \ 
$\forall i\in \mathbb Z \quad \theta^i = 
	\left\{
		\begin{array}{cc}
			\theta & \mbox{if $i$ is odd,} \\
			\theta^2 & \mbox{if $i$ is even.} 
		\end{array}
	\right.$
%\medskip

Let the generator matrix in polynomial form be 
\begin{equation}\label{eq:G(D)example}
G(D)=(1+\alpha D, \ \alpha +\alpha^2 D) = G_0 + G_1 D,
\end{equation}
where $G_0 = (1,\alpha)$ and  $G_1 = (\alpha,\alpha^2)$.
The generator matrix in scalar form (\ref{eq:G}) is 
\begin{equation}\label{eq:G_ex}
G=\left(
\begin{array}{ccccccc}
1\ \alpha  & \alpha \ \alpha^2        \\
           & 1\ \alpha^2       & \alpha^2 \ \alpha        \\
           &                   & 1\ \alpha         & \alpha \ \alpha^2        \\
           &                   &                   & 1\ \alpha^2       & \alpha^2 \ \alpha        \\
           &                   &   \dots                                        \\
\end{array}
\right).
\end{equation}
Here $\mu=1$, hence it is a \emph{unit memory} code.  The encoding rule is $v= uG$, or from (\ref{eq:enc_conv}) it is
\begin{equation}\label{eq:enc_rule_ex}
	v_t = u_t \theta^t (G_0) + u_{t-1} \theta^{t-1} (G_1),\ \mbox{for }t=0,1,\dots \ .
\end{equation}
From this example we can see that the class of skew convolutional codes \emph{extends}  the class of fixed  codes. Indeed, the codeword for the information sequence $u = 1,0,0,1$ is $v = (1,\alpha),(\alpha,\alpha^2), (0,0), (1,\alpha^2),(\alpha^2,\alpha)$, which cannot be obtained by any fixed $[2,1]$ memory $\mu=1$ code. 

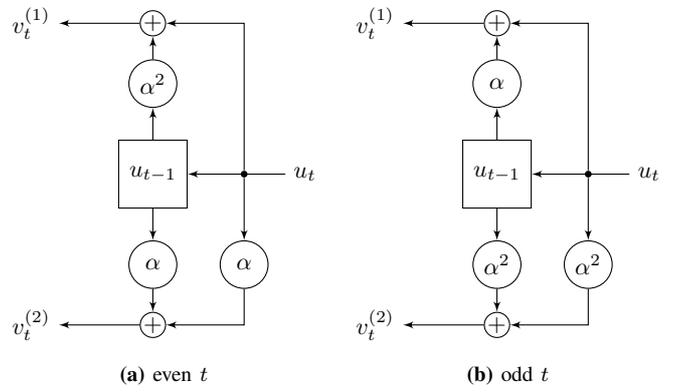
\begin{figure}[htp]
	\subfloat[][even $t$]{%
		\begin{tikzpicture}[scale=0.8, every node/.style={scale=0.9}]
		% square
		\draw (0,0) node[minimum size=1cm,draw](s) {$u_{t-1}$};
		% addition
		\draw (0,2.5) node(c1) [circ]{$+$};
		\draw (0,-2.5) node(c2) [circ]{$+$};
		% weight
		\node(w1) [draw,circle,minimum size=.7cm,inner sep=0pt] at (0,1.5) {$\alpha^2$};
		\node(w2) [draw,circle,minimum size=.7cm,inner sep=0pt] at (0,-1.5) {$\alpha$};
		\node(w3) [draw,circle,minimum size=.7cm,inner sep=0pt] at (1.5,-1.5) {$\alpha$};
		% dot
		\filldraw (1.5, 0) node(dot) [circle,fill,inner sep=1pt]{};
		% u,v
		\node(u) at (2.5,0) {$u_t$};
		\node(v1) at (-2,2.5) {$v^{(1)}_t$};
		\node(v2) at (-2,-2.5) {$v^{(2)}_t$};
		% arrows
		\draw [->,-latex'] (u) edge (s) (dot) edge [topath](c1) (c1) edge (v1);
		\draw [->,-latex'] (dot) edge (w3) (w3) edge [topath](c2) (c2) edge (v2);
		\draw [->,-latex'] (s) edge (w1) (w1) edge (c1);
		\draw [->,-latex'] (s) edge (w2) (w2) edge (c2);
		\end{tikzpicture}
	}
	\hfill
		\subfloat[][odd $t$]{%
		\begin{tikzpicture}[scale=0.8, every node/.style={scale=0.9}]
		% square
		\draw (0,0) node[minimum size=1cm,draw](s) {$u_{t-1}$};
		% addition
		\draw (0,2.5) node(c1) [circ]{$+$};
		\draw (0,-2.5) node(c2) [circ]{$+$};
		% weight
		\node(w1) [draw,circle,minimum size=.7cm,inner sep=0pt] at (0,1.5) {$\alpha$};
		\node(w2) [draw,circle,minimum size=.7cm,inner sep=0pt] at (0,-1.5) {$\alpha^2$};
		\node(w3) [draw,circle,minimum size=.7cm,inner sep=0pt] at (1.5,-1.5) {$\alpha^2$};
		% dot
		\filldraw (1.5, 0) node(dot) [circle,fill,inner sep=1pt]{};
		% u,v
		\node(u) at (2.5,0) {$u_t$};
		\node(v1) at (-2,2.5) {$v^{(1)}_t$};
		\node(v2) at (-2,-2.5) {$v^{(2)}_t$};
		% arrows
		\draw [->,-latex'] (u) edge (s) (dot) edge [topath](c1) (c1) edge (v1);
		\draw [->,-latex'] (dot) edge (w3) (w3) edge [topath](c2) (c2) edge (v2);
		\draw [->,-latex'] (s) edge (w1) (w1) edge (c1);
		\draw [->,-latex'] (s) edge (w2) (w2) edge (c2);
		\end{tikzpicture}
	}
	\caption{Encoder of the skew code $\C$.}\label{fig:enc_even}\label{fig:enc_odd}
\end{figure}

The  encoder (in controller canonical form \cite{JZ}) with generator matrix (\ref{eq:G(D)example}) is shown in Fig.~\ref{fig:enc_even}(a) for even $t$ and in Fig.~\ref{fig:enc_odd}(b) for odd $t$. The encoder has one shift register, since $k=1$. There is one $Q$-ary memory element in the shift register shown as a rectangular, where $Q= q^m =4$ is the order of the field. We need only one memory element since maximum degree of items in $G(D)$, which consists of a single row in our example, is $1$. A large circle means multiplication by the coefficient shown inside.

In general case of a $k\times n$ matrix $G(D)$, we define the degree $\nu_i$ of its $i$-th row as the
maximum degree of its components. The \emph{external degree} $\nu$ of $G(D)$ is the sum of its row degrees. The  encoder (in controller canonical form) of $G(D)$ over $\F_Q$ has $k$ shift registers, the $i$-th register has $\nu_i$ memory elements, and total number of $Q$-ary memory elements in the encoder is  $\nu$. 

For our example, the minimal code trellis, which has the minimum number of states, is shown in Fig.~\ref{fig:code_trellis}. The trellis consists of sections periodically repeated with period $\tau=m=2$. The trellis has $Q^\nu =4^1=4$ states labeled by elements of the field $\F_Q$. For the $t$-th section for time $t=0,1,\dots$, every edge connects the states $u_{t-1}$ and $u_t$ and is labeled by the code block $v_t$ computed according to the encoding rule (\ref{eq:enc_rule_ex}) as follows 
\begin{equation}\label{eq:trellis_labels}
v_t =
	\left\{
		\begin{array}{ll}
			u_{t-1} (\alpha,\alpha^2) + u_t (1,\alpha^2)   & \mbox{ for odd  } t, \\  
			u_{t-1} (\alpha^2,\alpha) + u_t (1,\alpha) & \mbox{ for even  } t.
		\end{array}
	\right.
\end{equation}
We assume that $u_{-1}=0$, i.e., the initial state of the shift register is $0$. 

\begin{figure}
	\centering
	\begin{tikzpicture}[scale=1.1, every node/.style={scale=0.8}]
	\node at (1,-.5){$t=0$};
	\node at (3,-.5){$t=1$};
	\node at (5,-.5){$t=2$};
	% trellis dots
	\foreach \x in {0,...,3} {
		\foreach \y in {0,...,3} {
			\filldraw (2*\x,\y) node [circle,fill,inner sep=2pt]{};
		}
	}
	\node[left] at (0,3){$0$};
	\node[left] at (0,2){$1$};
	\node[left] at (0,1){$\alpha$};
	\node[left] at (0,0){$\alpha^2$};
	% state (0) -- (0), (0) -- (1), (0) -- (a), (0) -- (a^2), t=0,2
	\foreach \x in {0,2} {
		\draw (2*\x,3) -- (2*\x+2,3) node [above,very near end] {$00$};
		\draw (2*\x,3) -- (2*\x+2,2) node [above,sloped,very near end] {$1 \alpha$};
		\draw (2*\x,3) -- (2*\x+2,1) node [above,sloped,very near end] {$\alpha \alpha^2$};
		\draw (2*\x,3) -- (2*\x+2,0) node [above,sloped,very near end] {$\alpha^2 1$};
	}
	% state (0) -- (0), (0) -- (1), (0) -- (a), (0) -- (a^2), t=1
	\foreach \x in {1} {
		\draw (2*\x,3) -- (2*\x+2,3) node [above,very near end] {$00$};
		\draw (2*\x,3) -- (2*\x+2,2) node [above,sloped,very near end] {$1 \alpha^2$};
		\draw (2*\x,3) -- (2*\x+2,1) node [above,sloped,very near end] {$\alpha 1$};
		\draw (2*\x,3) -- (2*\x+2,0) node [above,sloped,very near end] {$\alpha^2 \alpha$};
	}
	% state (1) -- (0), (1) -- (1), (1) -- (a), (1) -- (a^2), t=1,2
	% state (a) -- (0), (a) -- (1), (a) -- (a), (a) -- (a^2), t=1,2
	% state (a^2) -- (0), (a^2) -- (1), (a^2) -- (a), (a^2) -- (a^2), t=1,2
	\foreach \x in {1,...,2} {
		\foreach \y in {2,1,0} {
			\draw (2*\x,\y) -- (2*\x+2,3) ;
			\draw (2*\x,\y) -- (2*\x+2,2) ;
			\draw (2*\x,\y) -- (2*\x+2,1) ;
			\draw (2*\x,\y) -- (2*\x+2,0) ;
		}
	}
	% output of state (a^2) -- (a^2), t=1,2
	%        \node[below] at (3.8,0){$\alpha 0$};
	\draw (2*1,0) -- (2*1+2,0) node [below,very near end] {$\alpha 0$};
	%        \node[below] at (5.8,0){$1 0$};
	\draw (2*2,0) -- (2*2+2,0) node [below,very near end] {$1 0$};
	\end{tikzpicture}
	\caption{Time-varying minimal trellis of the skew  code $\C$.}\label{fig:code_trellis}
\end{figure}	
 
There are two important characteristics of a convolutional code: the free
distance $d_f$  and the slope $\sigma$ of increase of  the active burst distance, defined below as in \cite{JZ}.

The weight of a branch labeled by a vector $v_t$ is defined to be the   Hamming weight $w(v_t)$ of $v_t$. The weight of a path is the sum of its branch weights. A path in the trellis that  diverges from zero state, that does not use edges of weight $0$ from a zero state to another  zero state, and that returns to  zero state after $\ell$ edges is called a loop of length $\ell$ or \emph{$\ell$-loop}.   
 
 The \emph{$\ell$-th order active burst distance} $d_\ell^{\text{b}}$ is defined to be the minimum weight of $\ell$-loops in the minimal code trellis.   The slope is defined as $\sigma=\lim_{\ell \rightarrow \infty} d_\ell^{\text{b}}/\ell$. The free distance is $d_f = \min_{\ell} d_\ell^{\text{b}}$.
\begin{lemma}
	The skew convolutional code $\C$ defined by $G(D)$ in (\ref{eq:G(D)example}) has  the active burst distance
	$ d_\ell^{\text{b}} = \ell + 2$ for $\ell = 2,3,\dots$, the slope of the active distance  is $\sigma = 1$, and free distance is $d_f = 4$. 
\end{lemma}

General upper bounds on the free distance and the slope are given in \cite{PollaraAbdel}, which in our case of unit memory $[2,1]$ code $\C$ become  
$ d_{free} \le 2n - k + 1 = 4, \mbox{ and } \sigma \le  n-k =1.$
Hence, the skew code $\C$ defined by (\ref{eq:G(D)example}) reaches the upper bounds on $d_{free}$ and on the slope $\sigma$, hence, the \emph{code is optimal}. 

A generator matrix $G(D)$ of a skew convolutional code (and corresponding encoder)  is called \emph{catastrophic} if there exists an information sequence $u(D)$ of an infinite weight such that the code sequence $v(D) = u(D)G(D)$ has a finite weight.  The generator matrix $G(D)$ in (\ref{eq:G(D)example}) of skew convolutional code $\C$ with  $\theta = (\cdot)^q$  is non-catastrophic, since for $G(D)$ the slope $\sigma >0$. Note that in case of ordinary  convolutional code $\C'$, i.e., for $\theta = id$, the generator matrix  (\ref{eq:G(D)example}) is a catastrophic generator matrix of the repetition $[2,1]$ \emph{block} code with distance $d=2$.

A skew convolutional code, represented as  a $\tau$-periodic  $[n,k]$ code, can be considered as $[\tau n,\tau k]$ fixed code by \emph{$\tau$-blocking}, described in \cite{McE:1998}. The $[2,1]$ skew code $\C$ from our example has period $\tau=m=2$ and  can be written as $[4,2]$ \emph{fixed} code with generator matrix 
\begin{equation}\label{eq:G(D)_block}
G=\left(
\begin{array}{cccc}
1 & \alpha  & \alpha    & \alpha^2        \\
\alpha^2 D  & \alpha D  & 1   &  \alpha        \\
\end{array}
\right).
\end{equation}
In this way, known  methods to analyze fixed convolutional codes can be applied to skew convolutional codes.

\section{Dual codes}

Duality for skew convolutional codes can be defined in different ways. 

First, consider a skew convolutional code $\C$ over $\F$ in scalar form as a set of sequences as in ($\ref{eq:v}$). For two  sequences $v$ and $v'$, where at least one of them is finite, define the scalar product $(v,v')$ as the sum of products of corresponding components, where missing components are assumed to be zero.  We say that the sequences are orthogonal if $(v,v') = 0$. 
\begin{definition}\label{def:Cperp}
	The dual code $\C^\perp$ to a skew convolutional $[n,k]$ code $\C$ is an $[n,n-k]$ skew convolutional code $\C^\perp$ such that $(v,v^\perp)=0$ for all finite length words $v\in \C$ and $v^\perp \in \C^\perp$.  
\end{definition} 

Another way to define orthogonality is, for example, as follows. Consider two $n$-words $v(D)$ and $v^\perp(D)$ over $\Q^n$. We say that $v^\perp(D)$ is left-orthogonal to $v(D)$ if $v^\perp(D)v(D)=0$ and right-orthogonal if $v(D)v^\perp(D)=0$. A left dual code to a skew convolutional code $\C$ can be defined as 
\begin{equation*}
\C^\perp_\text{left} = \{v^\perp \in \Q^n : v^\perp(D)v(D)=0 \mbox{ for all } v\in \C\}. 
\end{equation*}
The dual code $\C^\perp_\text{left}$ is a left submodule of $\Q^n$, hence it is a skew convolutional code. 

We consider below dual codes according to Definition~\ref{def:Cperp}, since it is more interesting for  practical applications. Given a code $\C$ with generator matrix $G$, we show how to find a parity check matrix $H$ such that $G H^T=0$. 

Let a skew $[n,k]$ code $\C$ of memory $\mu$ be defined by  a polynomial generator matrix $G(D)$  in (\ref{eq:G(D)Ser}), which corresponds to the scalar generator matrix $G$ in (\ref{eq:G}). For the dual $[n,n-k]$ code $\C^\perp$ we write a transposed parity check matrix $H^T$ of memory $\mu^\perp$, similar to ordinary convolutional codes,  as 
\begin{equation}\label{eq:HT}
H^T=\left(
\begin{array}{clccccc}
H^T_0& H^T_1         &\dots & H^T_{\mu^\perp}&                         &      \\
         & \theta(H^T_0) &\dots &               &\theta(H^T_{\mu^\perp})      &           \\
%         &		             &\dots &               &\theta^2(H^T_{\mu'-1})&\theta^2(H^T_{\mu^\perp}) \\
         &                   &\dots                                    \\
\end{array}
\right),
\end{equation}
where $\text{rank} (H_0) = n-k$.
Similar to \cite{JZ}, we call the matrix $H^\perp$ the \emph{syndrome former} and write it in polynomial form as 
\begin{equation}\label{eq:HT(D)}
H^T(D) = H^T_0 + H^T_1 D + \dots + H^T_{\mu^\perp} D^{\mu^\perp}.
\end{equation} 
Then, we have the following \emph{parity check matrix} of the causal code $\C$ with the generator matrix (\ref{eq:G_equiv})
\begin{equation}\label{eq:H}
H=\left(
\begin{array}{clccccc}
H_0    \\
H_1       &   \theta(H_0)    \\
\vdots    & \vdots      \\
H_{\mu^\perp}  & \theta(H_{\mu^\perp-1})  &  \vdots  \\
          & \theta(H_{\mu^\perp}) \\
\end{array}
\right),
\end{equation}
which in case of $\theta = id$ coincides with the check matrix of an ordinary fixed convolutional code. 

From Definition~\ref{def:Cperp} we have that $vH^T=0$ for all sequences $v\in \C$ over $\F$. On the other hand, from (\ref{eq:enc1}) we have that every codeword $v(D)\in \C$ can be written as $v(D)=u(D)G(D)$. Hence, if we find an $n\times (n-k)$ matrix $H^T(D)$ over $\R$ of full rank such that $G(D)H^T(D) = 0$, then every codeword satisfies $v(D)H^T(D) = u(D)G(D)H^T(D)=0$ and vice versa, i.e., if  $v(D)H^T(D)=0$ then $v(D)$ is a codeword of $\C$. 
\begin{theorem}
	With the above notations, $G(D)H^T(D)=0$ if and only if $GH^T = 0$.
\end{theorem} 
  
We continue with the example given in Section~\ref{sec:example}. Let $H(D) = H_0 + H_1(D)$.  Using $G_0$ and $G_1$ from (\ref{eq:G(D)example}) and the condition  $G(D)H^T(D)=0$, we obtain $H_0 = (\alpha, 1)$ and $H_1 = (1,\alpha)$. Hence, $H(D) = (\alpha+D, 1+\alpha D)$ and 
\begin{equation*}\label{eq:HTex1}
H=\left(
\begin{array}{ccccccc}
\alpha \ 1      \\
1 \ \alpha & \alpha^2 \ 1      \\
           & 1 \ \alpha^2 & \alpha \ 1   & \vdots    \\
           &              &  1 \ \alpha        \\
%           &              &     \dots        
\end{array}
\right).
\end{equation*} 
Using $H$ one can draw the minimal trellis of the dual code $\C^\perp$ and  decode the original code symbol-wise  with the method in, e.g., \cite{Berkmann}. For high rate codes, this approach gains in
computational complexity as compared to the BCJR algorithm.

\section{Skew trellis codes}
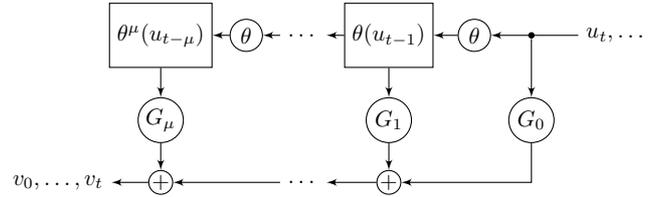
\begin{figure}%[t]
    \centering
	\begin{tikzpicture}[scale=0.75, every node/.style={scale=0.85}]
	% G-circles
	\node(w0) [draw,circle,minimum size=.7cm,inner sep=0pt] at (0,2.5) {$G_\mu$};
	%            \node(w3) [draw,circle,minimum size=1cm,inner sep=0pt] at (3,2) {$G_2$};
	\node(w4) [draw,circle,minimum size=.7cm,inner sep=0pt] at (4,2.5) {$G_1$};
	\node(w6) [draw,circle,minimum size=.7cm,inner sep=0pt] at (6.5,2.5) {$G_0$};
	% squares
	\draw (0,4)  node[minimum size=1cm,draw](s0) {$\theta^\mu(u_{t-\mu})$};
	%            \draw (3,4)  node[minimum size=1cm,draw](s3) {$\theta^\mu(u_{t-2})$};
	\draw (4,4)  node[minimum size=1cm,draw](s4) {$\theta(u_{t-1})$};
	% theta
	\node(th1) [draw,circle,minimum size=.5cm,inner sep=0pt] at (1.5,4) {$\theta$};
	%            \node(th4) [draw,circle,minimum size=.7cm,inner sep=0pt] at (4,4) {$\theta$};
	\node(th2) [draw,circle,minimum size=.5cm,inner sep=0pt] at (5.5,4) {$\theta$};
	% addition,arrows
	\foreach \x in {0,4} {
		\draw (\x,1.4) node(c\x) [circ]{$+$};
		\draw [->,-latex'] (s\x) edge (w\x) (w\x) edge (c\x);
	}
	% Input, Output
	\filldraw (6.5, 4) node(dot) [circle,fill,inner sep=1pt]{};
	\node(u) at (8,4) {$u_t, \dots$};
	\draw [->,-latex'] (u) edge (th2);
	\node(v) at (-1.8,1.4) {$v_0, \dots, v_t$};
	\draw [->,-latex'] (c0) edge (v);
	% dots, arrows
	\node(udots) at (2.5,4) {$\dots$};
	\node(vdots) at (2.5,1.4) {$\dots$};
	\draw [->,-latex'] (th2) edge (s4) (s4) edge (udots) (udots) edge (th1) (th1) edge (s0);
	%            \draw [->,-latex'] (c5) edge (c3) (c3) edge (vdots) (vdots) edge (c0);
	\draw [->,-latex'] (dot) edge (w6) (w6) edge [topath](c4) (c4) edge (vdots) (vdots) edge (c0);
	\end{tikzpicture}
	\caption{Encoder of a skew trellis code.}\label{fig:encoderTr}
\end{figure}

\begin{figure}[htp]
	\centering
	\begin{tikzpicture}[scale=0.8, every node/.style={scale=0.9}]
	% square
	\draw (0,0) node[minimum size=1cm,draw](s) {$\theta(u_{t-1})$};
	% theta
	\node(theta) [draw,circle,minimum size=.7cm,inner sep=0pt] at (1.5,0) {$\theta$};
	% addition
	\draw (0,2.5) node(c1) [circ]{$+$};
	\draw (0,-2.5) node(c2) [circ]{$+$};
	% weight
	\node(w1) [draw,circle,minimum size=.7cm,inner sep=0pt] at (0,1.5) {$\alpha$};
	\node(w2) [draw,circle,minimum size=.7cm,inner sep=0pt] at (0,-1.5) {$\alpha^2$};
	\node(w3) [draw,circle,minimum size=.7cm,inner sep=0pt] at (2.5,-1.5) {$\alpha$};
	% dot
	\filldraw (2.5, 0) node(dot) [circle,fill,inner sep=1pt]{};
	% u,v
	\node(u) at (3.5,0) {$u_t$};
	\node(v1) at (-2,2.5) {$v^{(1)}_t$};
	\node(v2) at (-2,-2.5) {$v^{(2)}_t$};
	% arrows
	\draw [->,-latex'] (u) edge (theta) (theta) edge (s) (dot) edge [topath](c1) (c1) edge (v1);
	\draw [->,-latex'] (dot) edge (w3) (w3) edge [topath](c2) (c2) edge (v2);
	\draw [->,-latex'] (s) edge (w1) (w1) edge (c1);
	\draw [->,-latex'] (s) edge (w2) (w2) edge (c2);
	\end{tikzpicture}
	\caption{Encoder of the skew trellis  code generated by  \eqref{eq:G(D)example}.}\label{fig5}
	\vspace*{-0.2cm}
\end{figure}
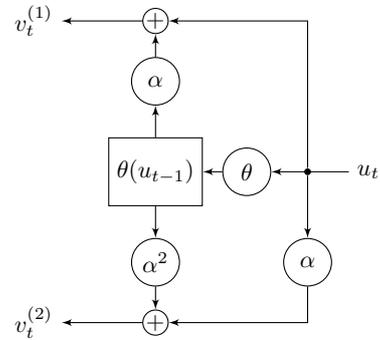

In this section,  by $\Q$ we denote the skew field of \emph{right} fractions of the ring $\R$ and we consider \emph{right} $\Q$-modules $\C$. Every module $\C$ is free  \cite[Theorem 1.4]{clark:2012}, i.e., it has a basis, and any two bases of $\C$ have the same cardinality, that is the dimension of $\C$. By $\F((D))$ we denote the skew field of \emph{right} skew Laurent series.
\begin{definition}[Skew trellis code]\label{def:SkewTrCode}
	A skew trellis $[n,k]$ code $\C$ over the field $\F$  is a right sub-module of dimension $k$ of the free module $\Q^n$.
\end{definition}
Every codeword $v(D)$ given by \eqref{eq:v(d)} can be written as 
\begin{equation}\label{eq:EncTr}
v^T(D) = G^T(D) u^T(D), 
\end{equation}
where $u(D)$ is an information word defined in \eqref{eq:u(d)} and $G(D)\in \Q^{k \times n}$ is a generator matrix of $\C$. Equivalently, one can consider a polynomial matrix $G(D)\in \R^{k \times n}$. Using \eqref{eq:u(D)ser} - \eqref{eq:G(D)Ser}, we rewrite the encoding rule \eqref{eq:EncTr} for sequences $u$ \eqref{eq:u} and $v$ \eqref{eq:v} over the field $\F_{q^m}$ as
\begin{equation}\label{eq:EncTr2}
v_t =  u_t G_0  + \theta(u_{t-1}) G_1 + \dots +  \theta^\mu (u_{t-\mu}) G_\mu.
\end{equation}
The corresponding encoders are  shown in Figs.~\ref{fig:encoderTr} and \ref{fig5} as  finite state machines. This allows us to obtain a code trellis and apply  known trellis-based decoding algorithms \cite{Vit}, \cite{BCJR}. The encoders can be obtained from the ones used for the ordinary convolutional code
 generated by $G(D)$  
 replacing the ordinary shift registers by  the skew shift registers introduced in \cite{SJB}. 
The case  $\theta = id$ gives  ordinary convolutional codes. For $\theta \ne id$ it follows from \eqref{eq:EncTr2} that the skew trellis code $\C=\{v\}$ as a set of sequences $v$ over $\F_{q^m}$ is  $\F_{q^m}$-nonlinear since so is the function $\theta(\cdot)$, but the code $\C$ is $\F_{q}$-linear.

%%%%%%%%%%%%%%%%%%%%%%%%%%%%%%%%%%%%%%%%%%%%%%%%%%%%

%\vspace 2pt
\section{Conclusion}
We defined  two new classes of skew codes over a finite field. The first class consists of linear skew convolutional codes, which are equivalent to  time-varying periodic convolutional codes but have as compact description as fixed convolutional codes. The second class consists of nonlinear trellis codes.

\bibliographystyle{IEEEtran}
\bibliography{IEEEabrv,refs}	

\end{document}